# Self-Supervised Speech Quality Assessment (S3QA): Leveraging Speech Foundation Models for a Scalable Speech Quality Metric

Mattson Ogg, Caitlyn Bishop, Han Yi, Sarah Robinson

*Johns Hopkins University Applied Physics Laboratory*, Laurel, USA
Correspondence: mattson.ogg@jhuapl.edu

***Abstract*— Methods for automatically assessing speech quality in real world environments are critical for developing robust human language technologies and assistive devices. Behavioral ratings provided by human raters (e.g., mean opinion scores; MOS) are considered the gold standard, but they are susceptible to variability between individual raters, cannot easily be generalized across corpora, and are labor-intensive to collect, thus limiting the acoustic challenges they can quantify. Here, we present a new, scalable method for automatically assessing speech quality: the self-supervised speech quality assessment (S3QA) model. First, we manipulated high quality utterances from multiple speech corpora, using a wide range of acoustic challenges intended to emulate common sources of quality degradation in the real-world: frequency filtering, reverberation, background noise, and digital compression. Second, we leveraged an existing, pre-trained speech foundation model, WavLM, to computationally derive a self-supervised training target that quantified speech degradation using the cosine distance between the clean and degraded versions of each utterance in the embedding space. Next, we trained a transformer-based model to predict these cosine distances, given only the degraded versions of the utterances. Finally, the trained model was evaluated on unseen test corpora of synthetic mixtures, NISQA, and VOiCES. We show that the S3QA model trained on this task accurately predicts degradation cosine distances across a wide range challenging acoustic conditions and is aligned with both behavioral ratings (MOS), speech technology performance (automatic speech recognition) and other important features of the held-out data (e.g., microphone distances). This model provides an automated, scalable method for assessing speech quality across a wide range of acoustic challenges. Our self-supervised training approach could be adapted to new use cases where acoustic simulations are accessible. The model is available online.[1]***

## I. Introduction

Human listeners can understand speech in everyday life despite a remarkable array of audio degradations, including frequency alterations (e.g., physical occlusion; virtual meetings; PA systems), reverberation, and background noise. However, this capacity begins to deteriorate with hearing loss, and assistive human language technology applications or devices often perform poorly in the face of real-world acoustic challenges. Quantifying speech quality *in-situ* across the wide variety of degradations a human or machine listener encounters could help recalibrate assistive devices, increase robustness in real-world scenarios, and improve algorithm development.

Speech quality has been a valuable metric for human language technology research, as well as for hearing assistance applications for many years [1]. Quality metrics have helped ensure reliable telecommunications [2], [3], supported the development of training and benchmark data for speech systems [4], [5], [6], guided speech enhancement research [7], [8], and accelerated speech synthesis methods [9], [10], especially when quality metrics are integrated into model training objectives [11], [12]. Measuring the perceptual consequences of degraded speech across a range of real-world conditions (e.g., noise [13], [14] or spectral degradation [15]) across different age groups is also critical for improving our understanding of hearing loss and for developing assistive technologies. Finally, degraded speech poses challenges for human language technologies like automatic speech recognition [16], [17] speaker recognition [18] and voice activity detection [19], [20].

Despite the impact of speech quality on many acoustic applications and perceptual tasks, assessing quality given the acoustic challenges of a real-world environment remains difficult. This difficulty arises because many assessment methods are domain-specific, and what quality is acceptable for a given task is often relative rather than absolute. Under controlled conditions, signal-to-noise ratio (SNR), signal-to-distortion ratio (SDR), or spectral power can be measured, but these do not relate directly to speech quality and are difficult to approximate from real-world recordings where the speech of interest might be mixed with background noise, reverberation (reverb) or other artifacts [21], [22]. Measures like PESQ [2] and STOI [3] have been useful for assessing telecommunications channels, but these require comparing target audio clips to a clean reference, which is often unavailable in real-world scenarios. Recently, deep learning methods have been developed to approximate PESQ and STOI without a clean reference [22], [23], [24], but the original metrics predicted by these models were still developed for specific telecommunications applications and may not generalize to other domains. Behavioral methods like mean opinion scoring (MOS [25]) or crowd-sourced voting [5] are adaptable to different acoustic challenges, and are the gold standard for speech quality evaluations. Neural network models can also be trained to predict these behavioral scores [22], [26], [27], [28]. However, eliciting responses from humans is a time-consuming and costly bottleneck that limits the scalability of ratings-based approaches and is incompatible with both real-time and large scale use cases where fine-tuning performance by hand is not feasible. Moreover, human raters often disagree with one another in their quality ratings [26] and the stimulus sets or paradigms used to gather these ratings may not be directly comparable, which limits generalizability across datasets [28].

[1] https://github.com/mogg64/S3QA

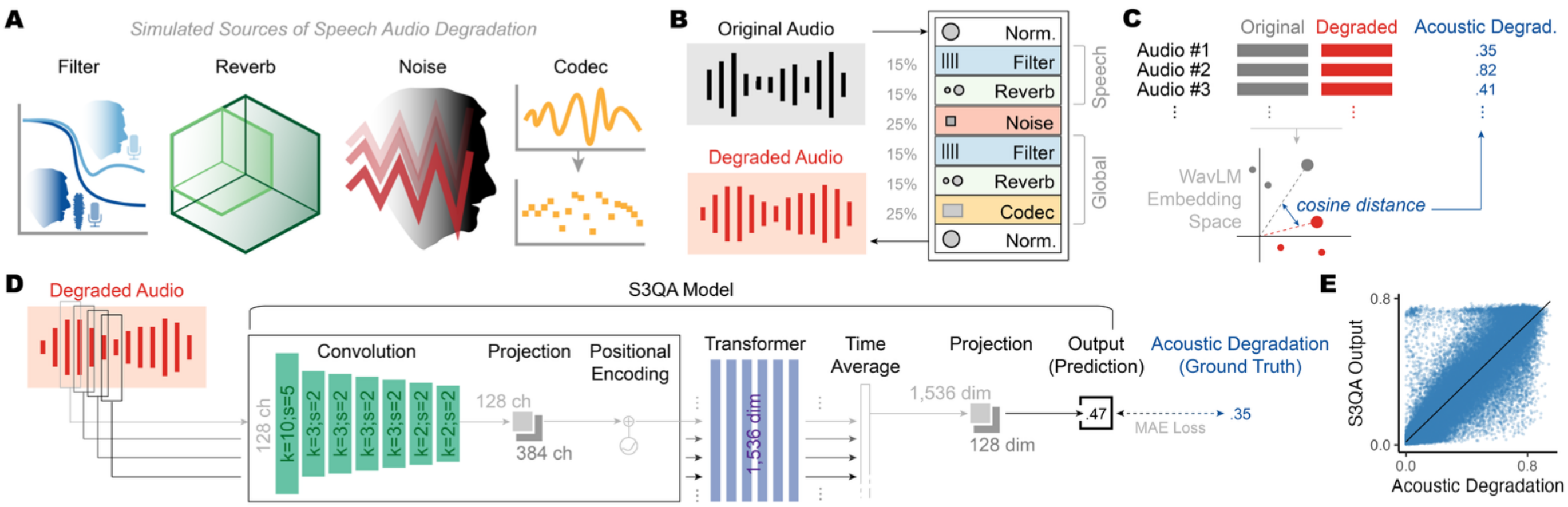

**Figure 1** Overview of our Self-Supervised Speech Quality Assessment (S3QA) approach along with a description of the model architecture, training and performance. A) Schematic of simulated acoustic degradations implemented for model training. B) Probabilities with which degradations were applied and the order in which they were applied. C) Method for obtaining the embedding distances from WavLM that were used to quantify degradation and to train the S3QA model. D) Diagram of information flow through the S3QA model architecture. E) Performance on our test partition for unseen English speech samples and background noises ($r_s$ = 0.91).

We observed a need for a general, scalable, and reference free speech quality measurement tool, suitable for multiple domains, including naturalistic far-field audio conditions, which can have unique signal degradations compared to other audio domains for which speech quality is often evaluated. For this study, multiple degradations (e.g., reverb, background noises) that impair real-world hearing and assistive devices were of interest. The multitude of real-world acoustic challenges complicates the elicitation of comprehensive behavioral ratings along all the relevant dimensions. Thus, we desired a computational solution that emulates human ratings, is theoretically motivated, avoids noise inherent in the labels used to train the model (c.f., behavioral, inter-rater noise in training labels for MOS-based approaches), is predictive of speech technology performance and has better precision for subtle differences in quality. To meet this challenge, we developed a self-supervised method for training a speech quality assessment model that relies on comparing embedding distances from WavLM, a pre-trained speech foundation model [30]. Embedding distances were calculated using high-quality speech utterances and degraded versions of the same utterances that underwent a wide array of acoustic augmentations. A transformer model was then trained to predict the WavLM embedding distances, given only the degraded utterances. The resulting self-supervised speech quality assessment (S3QA) model was evaluated on audio data from multiple domains, including a held-out dataset of unseen degraded audio, the NISQA test corpus [27], and the test partition of the VOiCES devkit [17].

Foundation models, such as WavLM, are ideal for generating self-supervision targets for this task because they encode rich features about speech that support performance on a wide range of downstream tasks (see Table I in [22]). Features from these models can also be used to predict the behavioral [31], [32] and neural [33], [34] responses of human listeners. This means that movement within the model's learned representational space (i.e., the embedding distances we used to train our speech quality model) likely corresponds to performance changes for a wide variety of speech technology tasks. Thus, we hypothesized that training our model to predict this distance measure could deliver a generalizable description of speech quality. We chose WavLM for this task because of its demonstrated downstream performance and its awareness of noise conditions from the auxiliary speech enhancement objectives used during training [30]. Other work has adapted speech recognition models for intelligibility prediction [35] or fine-tuned foundation models for MOS score prediction [29], but these approaches may be biased to a particular speech task (e.g., transcription) or involve the supervised learning of MOS scores, which are not ideal target labels for training due to subjective variability.

*Table I Data Used for Model Training and Evaluation*

| | Speech | RIR | Noise |
|---|---|---|---|
| **Training** | DAPS[e] (100/59,883) | BUT[^] (2,325/2,325) | FSD[%] (72,449/481,886) |
| Total: 3,964,700 | GLOBE[e] (572,159/880,391) | CD4M (468/468) | SONYC (13,538/94,766) |
| | LibriTTS[e] (149,736/1,735,984) | MIT (270/270) | US8k (8,732/7,333) |
| | CML-TTS[#] (987,749/1,196,555) | OpenAIR[$] (55/55) | WHAM (20,363/142,541) |
| | AISHELL-3 (63,262/91,887) | | AudioSet[+] (1,730,626/1,728,048) |
| **Validation** | GLOBE[e] (9,566/17,148) | RWCP[*] (143/143) | Freefield1010 (7,690/53,830) |
| Total: 400,198 | LibriTTS[e] (10,573/125,417) | | |
| | CML-TTS[#] (32,208/232,202) | | |
| | AISHELL-3 (24,773/25,431) | | |
| **Test** | VCTK (88,328/68,768) | AIR[*] (107/107) | ESC-50 (2,000/4,000) |
| Total: 101,989 | SHALCAS22A (14,580/33,221) | | |

Each corpus is listed along with the number of original audio files used and 4-second segments used in our train, validation, and test manifests

[+]Only the first 4-second segment of each file was used

[*] RWCP Type 1 and 2 only, from http://openslr.org/28/

[^] https://speech.fit.vutbr.cz/software/but-speech-fit-reverb-database

[$] https://www.openair.hosted.york.ac.uk

[%] FSD50k plus non-overlapping examples from noisy18k and Kaggle2019

[e] Included in English-only experiments

[#]CML-TTS langages: French, German, Italian, Polish, Portuguese, Spanish

We found that a transformer model trained using this approach could predict embedding distances very accurately on unseen test datasets ($r \geq 0.85$ in most cases). WavLM embedding distances and the outputs of our model were closely associated with speech transcription performance (i.e., edit distances between ASR transcripts for the clean and degraded utterances), and also with SNR and other degradation qualities such as microphone distances in the VOiCES corpus. Finally, even though our model wasn't trained to predict mean opinion scores, the model outputs correlate well with these ratings in the NISQA corpus.

## II. Methods

Figure 1 describes the overall approach for developing our self-supervised speech quality assessment model, including the degradation pipeline, the generation of training labels and model training.

### *A. Data and Pre-Processing*

We generated data for training, validation, and testing using high-quality corpora of public-domain speech [4], [6], [36], [37], [38], [39], [40], noises [41], [42], [43], [44], [45], [46], [47], [48], [49] (e.g., background noise, and other sound events, see Table I) and room impulse responses [50], [51], [52], [53]. When necessary, original recordings were converted to a single-channel and downsampled to 16 kHz. Speech files were trimmed of leading and trailing silences (defined as 30dB below the file's maximum root mean square (RMS) value using the librosa package [54]). Speech and noise recordings were then segmented into 4-second clips with a 1-second hop-size (except for AudioSet noises, from which only the first 4 seconds were extracted) and normalized to -35 dB LUFS (using the pyloudnorm [55] package). Any file shorter than 4 seconds was excluded. Table I summarizes the data used for the train, validation, and testing partitions.

Each clean speech segment underwent a randomized modification procedure similar to a data augmentation pipeline. All modifications were carried out using TorchAudio [56] functions unless otherwise specified. For each utterance, a set of degradations was sequentially applied based on a set of random draws with pre-defined likelihoods for each degradation:

- *Filter* (15% probability): If drawn, a butterworth filter was applied with parameters randomly selected among: 2nd- or 4th-order, high- or low-pass, with a cutoff between 10 and 3,500 Hz. Filters were applied using TorchAudio's *filtfilt()* function.
- *Room Impulse Response* (15% probability): If drawn, a room impulse was randomly selected from the set of impulses corresponding to the appropriate partition and applied to the segment using the *fftconvolve()* function.
- *Background Noise* (25% probability): If drawn, a random 4-second background noise was selected from the set of noise clips corresponding to the appropriate partition and applied using the *add_noise()* function with an SNR randomly selected as an integer between -30 and 30 dB.
- *Filter* (15% probability): Same as above. Could be applied even if a filter was previously applied (with parameters drawn independently of the first filter).

- *Room Impulse Response* (15% probability): Same as above, but only applied if a room impulse response was *not* previously applied.
- *Codec* (25% probability): If drawn, one of three codecs was randomly selected to be applied (mp3, ogg-vorbis, or GSM). If GSM was selected, the audio was first downsampled to 8 kHz, then the codec was applied using the *apply_codec()* function and the audio was then resampled to 16 kHz. Degradations using mp3 or ogg-vorbis were applied by writing a temporary file to disk using TorchAudio's *save()* function with the application of a specified degradation randomly drawn (-1 to 10 for ogg-vorbis, and a list of 5 to 20, 30, 40, 65, 85, 100, 115, 130, 190, 320 for mp3 which determine the file size and bitrate of the output).

Figure 1 describes the data modification pipeline to create degraded audio clips for each clean audio clip. Modified files were normalized to -35 dB LUFS. Randomized modifications were performed four times for each 4-second utterance to build each partition.

After modified versions of the utterances were generated a number of metrics were calculated for training and evaluation of the S3QA model. First, we extracted WavLM-Large model embeddings for each clean and modified utterance, averaging the representation from the last layer of the model over time (features are output by the WavLM model roughly every 20 milliseconds [30], [57]), and then we calculated the cosine distance between the embeddings corresponding to the clean and modified versions. WavLM models used weights that had been pre-trained but these were *not* fine-tuned for any specific downstream task [2]. The maximum cosine distance of the utterances in the training set (1.18) was used as a linear scaling factor to scale target distances to between zero and one during training to simplify the interpretability of model outputs and to homogenize training for different ablation experiments. This same scaling factor (derived from the train partition) is applied to WavLM embedding distances in the validation and test partitions to evaluate the mean absolute error of the model. We also transcribed all utterances using Whisper ("large-v3"[58][3]), and we calculated the edit distance between the transcriptions of the clean and modified utterances at the word and character-level; specifically, we derived word error rate (WER) and character error rate (CER) using the clean utterance's transcript as the reference and the modified utterance's transcript as the prediction (however, per the standard convention we only evaluate the Mandarin SHALCAS22A corpus using CER; e.g., [59]). We also calculated SI-SDR, PESQ, and STOI for each clean and modified utterance pair using torchmetrics [60].

### B. S3QA Model Architecture and Training

We trained our S3QA transformer model to predict the WavLM distances derived for each clean and modified utterance pair, using only the modified, degraded utterance as input (i.e., in a reference-free set up). The S3QA architecture is a smaller implementation of other models that learn representations of data directly from the time-domain (e.g., [30], [57], [61])

Our model takes a degraded utterance as input, and directly processes the signal in the time-domain using a front-end made up of a series of seven 1-dimensional convolutional layers (each with 128 channels, strides = 5,2,2,2,2,2,2 and kernel widths = 10,3,3,3,3,2,2 corresponding to layers 1 to 7, respectively). The 1-dimensional convolution in each of these layers is followed by a layer-norm and gelu activation function. After this series of convolutional layers, the data are projected from 128 to 384 followed by a gelu activation layer and positional encoding. This is then passed to six transformer encoder layers (each with 8 attention heads, a 1536 inner feed forward dimension, gelu activation and 5% layer drop probability). The output of the last layer of the transformer is averaged over the time dimension and is projected to a 128-dimension embedding layer (with gelu activation function) before reaching the output layer.

The model is trained to minimize a mean-squared-error loss function to predict the WavLM distance between each modified utterance and its clean reference. Values output by the model are best thought of as "degradation scores," where a higher value indicates a more degraded speech sample. The model is trained for 40 epochs using the Adam optimizer. The learning rate is scaled from 1e-05 to 5e-04 over the first 15 training epochs, and then scaled down to 5e-09 over the next 25 epochs. During training, the duration of each input batch is randomly truncated to between 1 to 4 seconds to make the model robust to different-length inputs. At the end of training, we retain the model weights from the epoch with the lowest loss on the validation partition. Training was carried out on an Nvidia A100 GPU with 16 workers, using a batch size of 128.

[2] via: https://github.com/microsoft/unilm/tree/master/wavlm
[3] via: https://github.com/openai/whisper

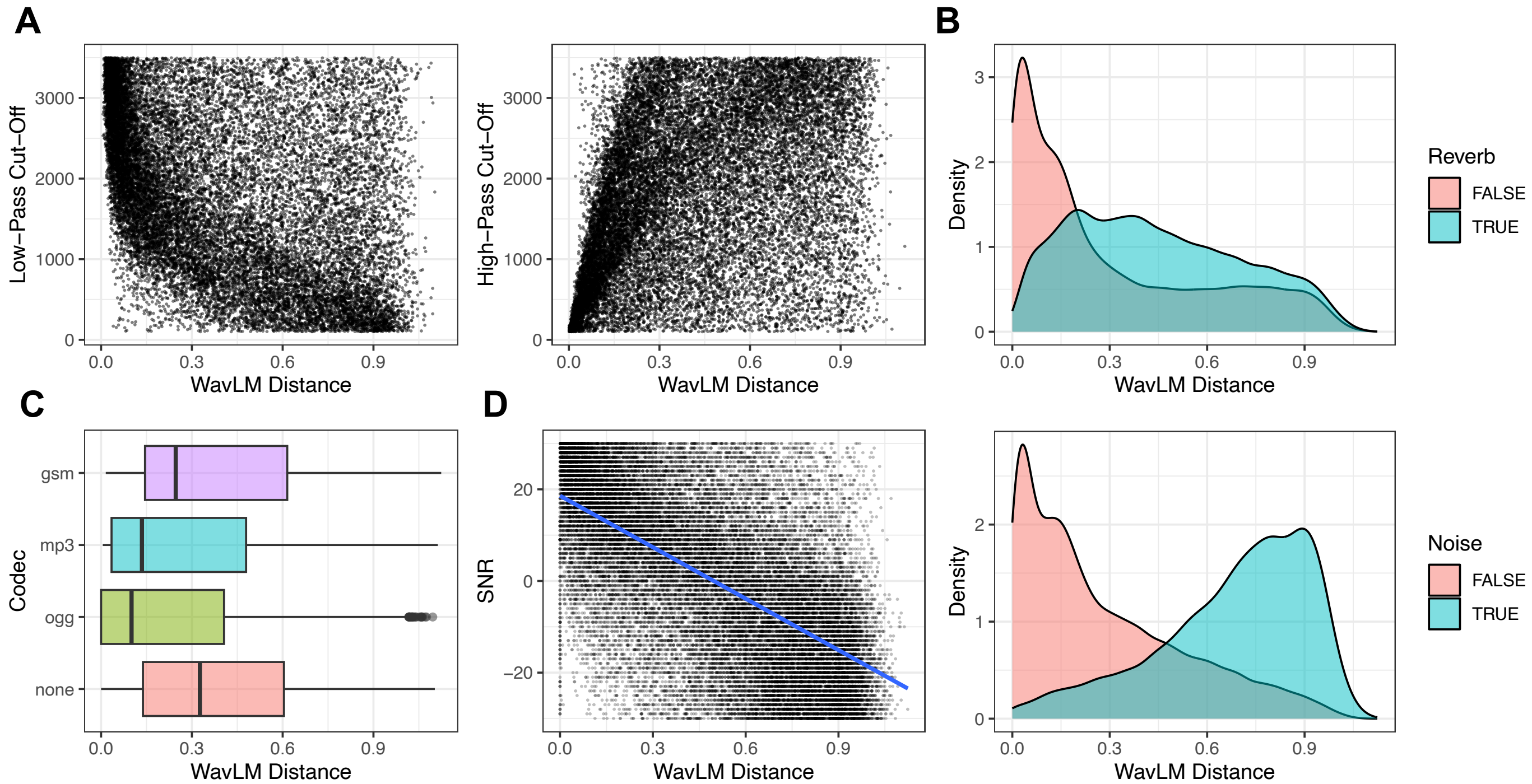

**Figure 2** The relationship between WavLM embedding distances and different acoustic manipulations in the test partition we generated. Note, individual samples may contain multiple acoustic manipulations in addition to the specific manipulation considered. A) High- and low-pass filter cut-off frequencies for speech segments where these filters were applied. B) Application (or lack thereof) of convolutional reverb. C) The application of different codecs (in the presence of other degradations). D) The influence of background noise on WavLM distances as a function of SNR, and the overall shift in embedding distance relative to SNR (left) and the overall application (or lack thereof) of background noise (right), the blue line indicates a linear fit to the data.

### C. *Performance Evaluation and Baseline Comparisons*

We compared the performance of our model to two open-source MOS-prediction models (NISQA [27] and the TorchAudio-Squim [22] implementation of NORESQA-MOS [28]). We evaluated these models on our held-out test corpus. We also evaluated these baseline models and our S3QA model on the combined test partition of the NISQA corpus [27], for which human-rated MOS scores are available, and on the test partition of the VOiCES devkit corpus [17], which contains controlled far-field audio recordings and their corresponding metadata. We chose these corpora specifically because they contained clean-reference samples, and limited our analyses of these external datasets to utterances where a clean reference was available so that we could derive the full set of objective metrics for evaluation (i.e., "ground truth" WavLM embedding distances, WER and CER edit distances, SI-SDR, PESQ, and STOI). This provides quantitative assessments of the utility of the S3QA model's degradation score (e.g., impact to downstream tasks such as transcription). The accuracy with which S3QA predicts WavLM embedding cosine distances is assessed using mean-absolute-error (MAE) and Spearman correlations between the model's outputs and the target WavLM cosine distances. Performance across models, objective metrics and acoustic features is carried out using Spearman correlations with a Bonferroni correction for multiple comparisons (for the number of correlations run for each dataset).

## III. Results

### A. *Association Between Target WavLM Distances and Acoustic Degradations*

Foundation model embeddings track critical features of speech [62], [63], [64], but their relationship to different acoustic degradations that impair downstream model or human-listener performance is, to our knowledge, not well characterized. In order for WavLM distances between clean and degraded utterances to be a useful self-supervision signal for speech quality assessment, we wanted to first measure the association between degradations imposed by our data generation pipeline and movement within WavLM's representational space by examining a test partition created using this augmentation pipeline.

The influence of different acoustic degradations on WavLM embedding distances relative to clean reference audio is shown in Figure 2. Low-pass filtering significantly increased WavLM embedding cosine distances (0.29 to 0.34 on average, Wilcoxon rank sum test, $W$ = 743898510, $p <$ 0.001) as did high-pass filtering (0.29 to 0.34 on average, $W$ = 704430836, $p <$ 0.001), and these shifts were significantly associated with the filter's cut-off frequency ($r_s$ = -0.47, $p <$ 0.001 and $r_s$ = 0.44, $p <$ 0.001 for low- and high-pass filter cut-offs respectively). Reverb also had a significant impact on WavLM embedding distances (0.25 to 0.38 on average, $W$ = 780237336, $p < 0.001$). Background noise had the strongest impact on embedding distances (0.24 to 0.57 on average, $W$ = 204933466, $p < 0.001$), and there was a strong association between SNR and embedding distance ($r_s$ = -0.64, $p < 0.001$). Finally, the application of each codec also had a significant influence on WavLM embedding distances (Kruskal-Wallis test, $\chi^2$ = 6642, $p < 0.001$), but mostly these *reduced* embedding distances (ogg-vorbis: 0.32 to 0.19 on average, $W$ = 543975188, $p < 0.001$; mp3: 0.32 to 0.23 on average, $W$ = 506999852, $p < 0.001$) relative to segments with no codec applied (except for GSM: 0.32 to 0.32 W = 402123152, $p$ = 0.94). This may seem counterintuitive, but recall that the goal of an audio codec is usually to ensure transmission and storage of important speech or acoustic features, so it is reasonable that they would increase the saliency of speech information (and thus reduce WavLM distances) in the presence of other acoustic degradations. We still include these acoustic manipulations, because they subtly alter the acoustic information in the audio signals and we want the resulting speech quality model to be aware of codecs.

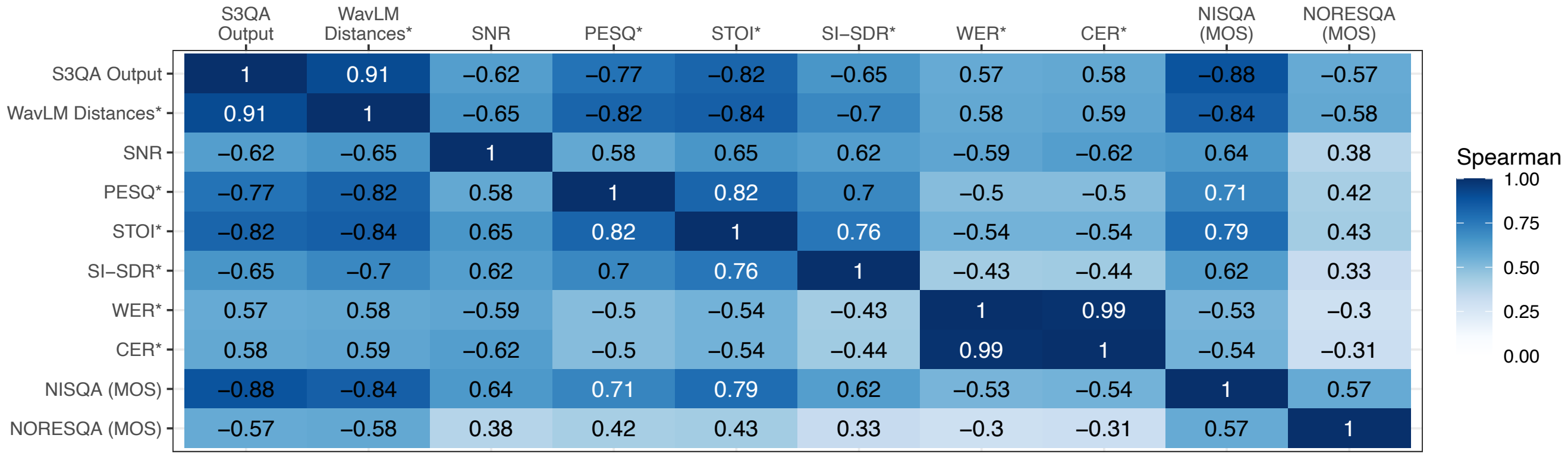

**Figure 3** Model performance and Spearman correlation among metrics for the VCTK test partition. All tests $p < 0.05$ following Bonferroni correction for multiple comparisons. Measures with asterisks indicate that they are calculated using a matched reference. Cells are colored based on the absolute value of each correlation (since some metrics vary in opposite directions) to indicate overall correlation strength. The signed correlation value is provided within each cell.

Together, these results demonstrate that WavLM embeddings are sensitive to the kinds of degradations we used to train our S3QA model, and could thus serve as a self-supervision signal (i.e., pseudo-label) during model training.

### *B. S3QA Performance: Test Partition*

The S3QA model learned to predict WavLM distances and generalized well to unseen data. The model is able to accurately predict WavLM distances for mixtures generated from unseen datasets in our test partition ($r_s$ = 0.88, $p < 0.001$, $MAE$ = 0.08). Overall, however, the model appeared reluctant to output more extreme degradation scores (e.g., few scores over 0.75). Finally, to ensure the validity of our approach, we also evaluated the clean (unmodified) utterances using the S3QA, and found that the degradation scores output by the model were very low (inter quartile range (IQR): 0.01 to 0.02), as expected.

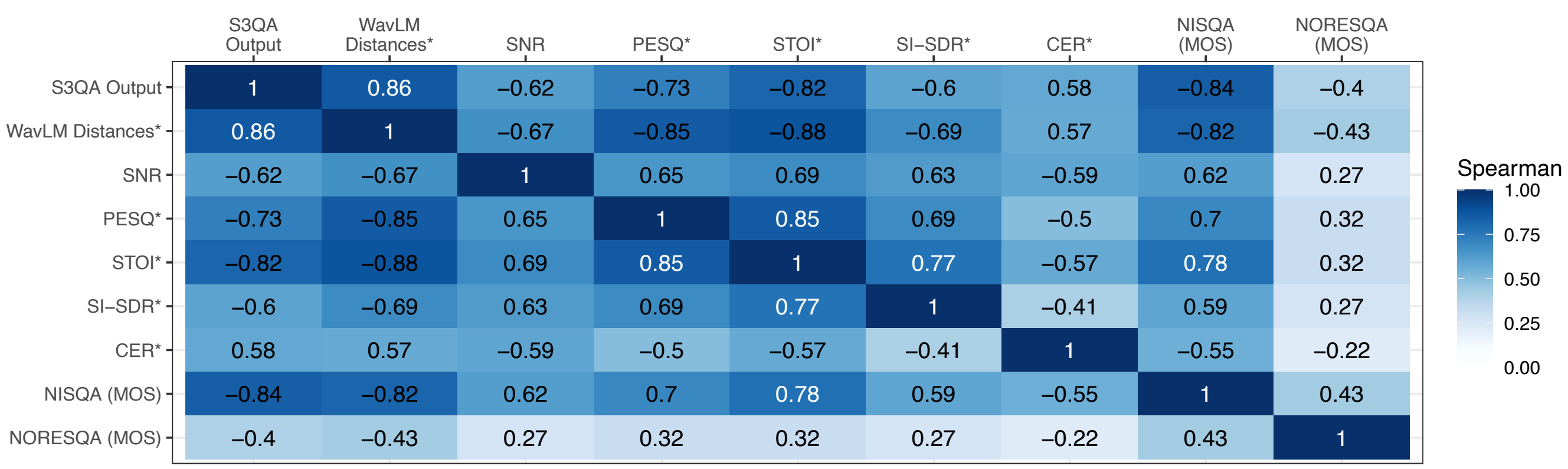

**Figure 4** Model performance and Spearman correlation among metrics for the SHALCAS22A test partition. All tests $p < 0.05$ following Bonferroni correction for multiple comparisons. Measures with asterisks indicate that they are calculated using a matched reference. Cells are colored based on the absolute value of each correlation (since some metrics vary in opposite directions) to indicate overall correlation strength. The signed correlation value is provided within each cell.

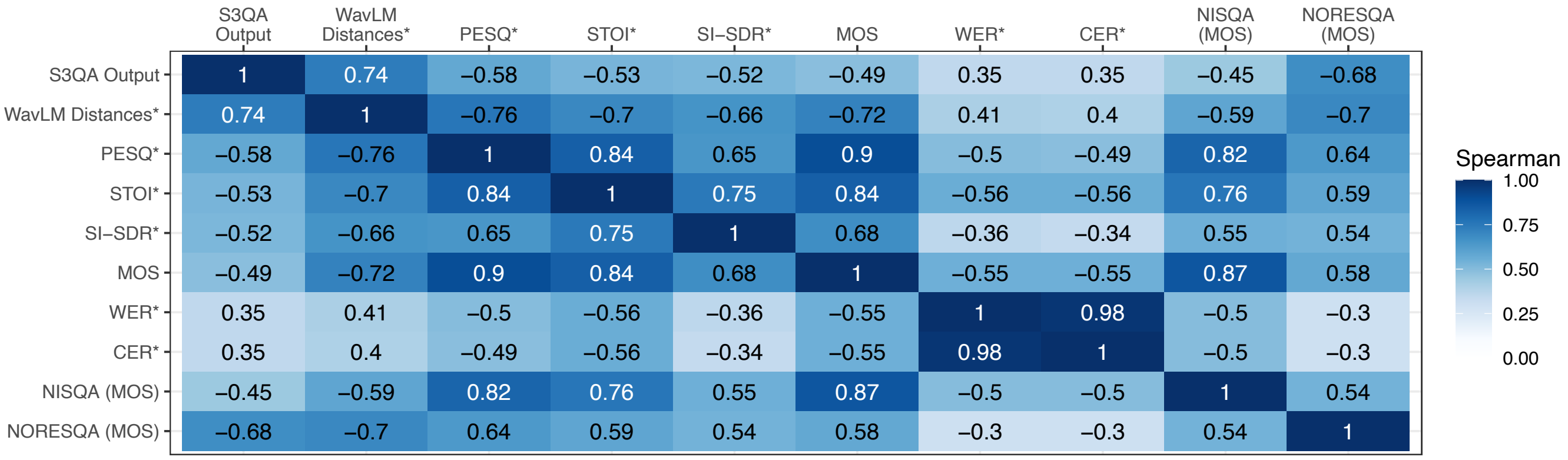

**Figure 5** Model performance and Spearman correlation among metrics for the NISQA corpus. All tests $p < 0.05$ following Bonferroni correction for multiple comparisons. Measures with asterisks indicate that they are calculated using a matched reference. Cells are colored based on the absolute value of each correlation (since some metrics vary in opposite directions) to indicate overall correlation strength. The signed correlation value is provided within each cell.

Follow-up analyses (summarized in Figure 3 and Figure 4) indicated that performance was slightly better on held-out English speech (from the VCTK dataset, $r_s$ = 0.91, all Bonferroni-corrected $p < 0.001$, $MAE$ = 0.08) than on Mandarin Chinese speech (from the SHALCAS22A dataset, $r_s$ = 0.86, $p < 0.001$, $MAE$ = 0.09). This performance gap between languages could follow from WavLM having been trained only on English data, even though S3QA was trained on WavLM distances for speech from multiple languages. Overall, this bias might result in sub-optimal performance for non-English speech, although we note that performance for Mandarin Chinese is still near ceiling. This issue is explored in follow-up analyses below.

S3QA distance predictions were also significantly correlated with WER and CER edit distances between clean and degraded utterances for English (on VCTK, correlation between S3QA degradation score outputs and word-level edit distances: $r_s$ = 0.57, $p < 0.001$) and Mandarin Chinese (on SHALCAS22A, correlation between S3QA degradation score output and character-level edit distances: $r_s$ = 0.58, $p < 0.001$). The association between S3QA degradation scores and edit-distances was higher than any other model's output or metric, though performance was comparable with objective metrics such as SI-SDR and STOI (see Figure 3 and Figure 4).

### C. *Ablation Studies*

We tested different formulations of the S3QA model architecture and training procedure to explore possible optimizations to performance and efficiency.

First, we explored using cosine distances from other embedding models as training targets. We used the Boot Strap Your Own Latent approach for training a self-supervised CNN model of audio data (BYOL-A [65]) on ten thousand hours of speech and non-speech data (full model training described in [64], this model also did not pre-train on the datasets used in our test partition). While a new S3QA transformer model was able to very accurately learn these BYOL-A model distances ($r_s$ = 0.93, $p < 0.001$, $MAE$ = 0.05 for the test partition overall), additional analyses indicated that the S3QA model outputs trained to predict BYOL-A distances were not as strongly related to other objective metrics and downstream human language technology capabilities (lower correlation with WER and CER edit distances in English on VCTK, $r_s$ = 0.48, $p < 0.001$, and in Mandarin Chinese for SHALCAS22A, $r_s$ = 0.52, $p < 0.001$), as compared with the WavLM distances.

Next, given that WavLM was pre-trained on English data only, and performance was slightly lower on degraded SHALCAS22A speech samples, it could be possible that including non-English speech data in S3QA training might have hurt performance due to a domain shift introduced into the WavLM's representational space from non-English training examples. Thus, we trained another S3QA model using only English speech samples. This model obtained similar overall performance to the original S3QA model for the test partition ($r_s$ = 0.88, $p < 0.001$, $MAE$ = 0.09), but a much higher error rate for SHALCAS22A data in the test partition (MAE = 0.13); correlation with edit distances was comparable to the original model (for both VCTK, $r_s$ = 0.58, $p < 0.001$, and SHALCAS22A, $r_s$ = 0.57, $p < 0.001$). Therefore, we conclude that including multiple languages in S3QA training leads to improved generalizability without negatively impacting performance.

Finally, we examined whether increasing the number of model parameters in the S3QA architecture might improve performance. We trained another S3QA model (on the original English and non-English training partition) with more CNN filters in the front-end (256 projected to a 480-dimension linear layer), and a larger transformer (10 heads, 8 transformer encoder layers, and a feedforward dimension of 1920). However, this model only produced marginally better results ($r_s$ = 0.89, $p < 0.001$, $MAE$ = 0.08 for the test partition overall; $r_s$ = 0.91, $p < 0.001$, $MAE$ = 0.07 for VCTK samples in the test partition; $r_s$ = 0.87, $p < 0.001$, $MAE$ = 0.10 for SHALCAS22A samples in the test partition), despite nearly doubling the model size (from 14.7 million parameters to 28.6 million).

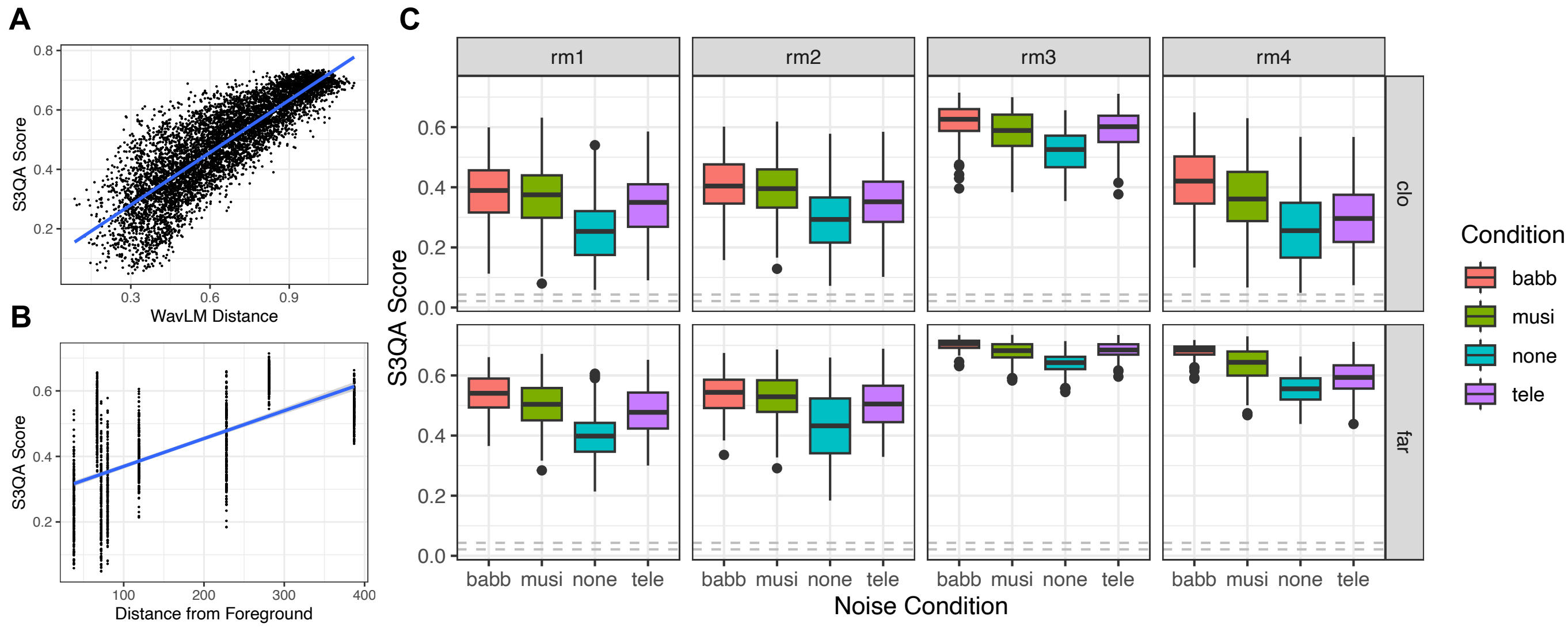

**Figure 6** S3QA performance on the VOiCES devkit test corpus. S3QA scores on these data are presented in relation to different room, recording, and interference conditions. A) Target WavLM distances. B) Physical distances between the microphone and foreground speech source. C) S3QA degradation scores as a function of the different room and background noise conditions in VOiCES.

These analyses did not yield compelling evidence to alter the design of S3QA; therefore we proceeded with evaluating the original S3QA model using additional unseen audio.

*D. S3QA Performance: External Datasets*

Two additional held out datasets were identified to support evaluation of our S3QA model. The developers of the NISQA model released their training, validation, and testing data [27]. Since NISQA was originally designed to predict MOS scores, this dataset allows us to examine how S3QA outputs align with behavioral ratings of speech quality. This dataset is also commonly used as a benchmark in studies of speech quality evaluation. We obtained the NISQA test data and assessed all samples with an available clean reference audio clip, obtaining the same panel of objective metrics and also considering the human-labelled MOS scores for each degraded utterance.

S3QA predicted WavLM embedding distances in the NISQA test corpus less accurately than for our internal test corpus ($r_s$ = 0.74, all Bonferroni-corrected $p < 0.001$, *MAE* = 0.11; see Figure 5), however, embedding distances were still significantly associated with word-level edit distances between Whisper-generated transcripts for the clean and degraded utterances ($r_s$ = 0.35, $p < 0.001$). Finally, while S3QA was not explicitly trained to predict MOS scores, its embedding-distance-based degradation scores were still significantly associated with MOS behavioral ratings ($r_s$ = -0.49, $p < 0.001$).

We also compared S3QA performance on the NISQA dataset with open-source systems NISQA and NORESQA-MOS. The NISQA model performed well on this test dataset (correlation with MOS scores: $r_s$ = 0.87, all Bonferroni-corrected $p < 0.001$ correlation with word-level-edit distances: $r_s$ = -0.50, $p < 0.001$), as expected due to the domain match between training and test audio. NORESQA-MOS (which was not previously trained or evaluated on the NISQA test corpus, to our knowledge) performance was comparable to S3QA, correlating better with MOS scores, but worse with edit-distances (correlation with MOS scores: $r_s$ = 0.58, $p < 0.001$; correlation with word-level-edit distances: $r_s$ = -0.30, $p < 0.001$).

Finally, we noted that the NISQA corpus overall appeared to be less degraded and was subject to less diverse kinds of audio signal interference than our original training corpus, which targeted more far-field audio conditions. For example, our internal test corpus differed from the NISQA data in terms of numerous objective measures (SI-SDR, -6.03 vs 6.92 on average, $W$ = 9562956, $p < 0.001$, SI-SNR, -5.88 vs 7.00 on average, $W$ = 9610660, $p < 0.001$, and word-level-edit distances, 0.41 vs 0.10 on average, $W$ = 26975413, $p < 0.001$). This could mean S3QA is better suited for characterizing more distant and noisier recordings; based on its training, it may lack the ability to finely discriminate among milder signal degradations.

To test this hypothesis, and to evaluate SQ3A on data that is known to have more severe degradations, we also evaluated the VOiCES devkit dataset. This dataset comprises utterances from LibriSpeech [66] played back in multiple rooms, often in the presence of complex background sounds. No behavioral scores exist for these data, but other extensive documentation allows us to understand how the S3QA scores are influenced by different kinds of interference.

S3QA performance on the VOiCES devkit is reported in Figure 6 (see Figure 7 for comprehensive comparison across metrics). S3QA predicted the target WavLM embedding distances better than for the NISQA corpus, and comparable with our internal test corpus ($r_s$ = 0.87, all Bonferroni-corrected $p < 0.001$, *MAE* = 0.10), and were significantly associated with word-level edit distances between Whisper-generated transcripts for the clean and degraded utterances ($r_s = 0.53$, $p < 0.001$). This latter result is likely due to the more challenging data in VOiCES compared to NISQA that could have caused Whisper to produce a wider range of errors for the far-field utterances. S3QA scores were also significantly correlated with the reported distance between each microphone and the foreground speech source ($r_s = 0.55$, $p < 0.001$, run only over utterances with no background noise) and were significantly influenced by the different rooms used for recording (Kruskal-Wallis test, $\chi^2 = 2079.1$, $p < 0.001$) and background noise conditions (Kruskal-Wallis test, $\chi^2 = 411.8$, $p < 0.001$). On these data S3QA also correlated more closely with key outcomes like word-level-edit distances and microphone distance from foreground sound sources than both NISQA and NORESQA-MOS (Figure 7).

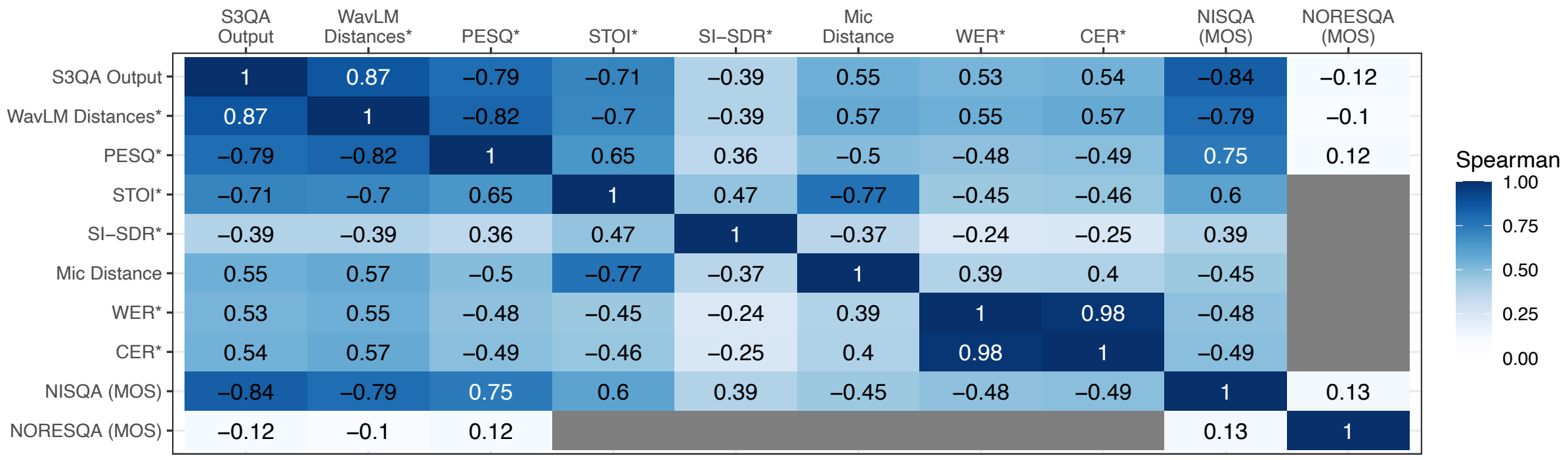

**Figure 7** Model performance and Spearman correlation among metrics for the VOiCES devkit corpus. All tests $p < 0.05$ following Bonferroni correction for multiple comparisons and tests that do not survive multiple comparison correction are greyed out. Measures with asterisks indicate that they are calculated using a matched reference. Cells are colored based on the absolute value of each correlation (since some metrics vary in opposite directions) to indicate overall correlation strength. The signed correlation value is provided within each cell.

## IV. Discussion

Our S3QA model can automatically assess speech quality after being trained to predict speech foundation model embedding distances between clean references and degraded utterances. This embedding-distance-based approach provides a compelling alternative to attempting to estimate behavioral ratings, which are labor-intensive and have poor resolution, or downstream speech-technology tool performance, which is an indirect measure of speech quality. We show that S3QA, trained to predict these distances from degraded utterances, can perform well on unseen data, and that the model's degradation scores (embedding distance estimates) are correlated with many outcomes related to speech quality including downstream speech-technology performance (quantified here by edit distances between the clean and degraded utterances), and human behavioral judgments (MOS scores).

Using a behavioral methods (e.g., MOS scoring) to search the space of acoustic manipulations we studied here would require significant time and resources. Alternatively, comparing foundation model distances is simple and scalable. Our choice to rely on foundation model embeddings may have biased our output towards quality scores relevant to machine, rather than human, perception, reflected in the correlations between S3QA degradation scores and edit-distances for automated transcripts. However, the high performance of audio foundation models has recently been closing the gap between human and machine perception for many tasks, and also appears to align closely with neural processing of acoustic information [32], [33], [34]. Indeed, our S3QA model outputs were correlated with behavioral assessments in the NISQA corpus, which suggests that machine and human perception (at least for high-level judgments of speech quality), have significant overlap.

Compared to a behavioral training target, a computational training target derived from a speech foundation model may also be favorable given benefits in scalability and the consistency of these scores relative to differences in MOS behavioral ratings across different listeners and test datasets. Moreover, WavLM distances (and thus the S3QA model's output) have better resolution than a given listener's MOS ratings (i.e., continuous values compared to discrete integers from 1 to 5).

Finally, the foundation models that guided our approach learn features that can perform a wide range of downstream speech tasks [30], making them suitable to supporting a general-purpose quality metric. Thus, the emphasis placed by S3QA on machine perception might also be advantageous for many applications, such as for integration with speech enhancement [7], [8] or speech synthesis [9], [10] systems. These tasks require preserving what is being said as well as the characteristics of how the utterance is delivered and who is speaking. These linguistic and paralinguistic features as indexed by audio foundation models could be captured n S3QA scores as an objective function to improve the performance of future audio generation applications.

## Acknowledgment

The authors acknowledge support from Independent Research and Development (IRAD) funding of the Johns Hopkins University Applied Physics Laboratory. Research reported in this paper was also supported by the NIH awards UH3NS136631, R24MH136632, R24DA064430, and U24MH136628. Feedback from ChatGPT was used for assistance with copy editing while drafting this report.